\documentclass[conference]{IEEEtran}
\IEEEoverridecommandlockouts
\usepackage{cite}
\usepackage{amsmath,amssymb,amsfonts}
\usepackage{algorithm}
\usepackage{algorithmic}
\usepackage{graphicx}
\usepackage{subfigure}
\usepackage{textcomp}
\usepackage{xcolor}
\usepackage{footnote}
\ifCLASSINFOpdf
\else
\fi
\begin{document}

\onecolumn
\copyright 2019 IEEE.  Personal use of this material is permitted.  Permission from IEEE must be obtained for all other uses, in any current or future media, including reprinting/republishing this material for advertising or promotional purposes, creating new collective works, for resale or redistribution to servers or lists, or reuse of any copyrighted component of this work in other works.

\twocolumn
\title{$\pi$-BA: Bundle Adjustment Acceleration on Embedded FPGAs with Co-observation Optimization}


\author{\IEEEauthorblockN{Shuzhen Qin}
\and
\IEEEauthorblockN{Qiang Liu}
\and
\IEEEauthorblockN{Bo Yu}
\and
\IEEEauthorblockN{Shaoshan Liu}}


\maketitle

\begin{abstract}
Bundle adjustment (BA) is a fundamental optimization technique used in many crucial applications, including 3D scene reconstruction, robotic localization, camera calibration, autonomous driving, space exploration, street view map generation etc. Essentially, BA is a joint non-linear optimization problem, and one which can consume a significant amount of time and power, especially for large optimization problems. Previous approaches of optimizing BA performance heavily rely on parallel processing or distributed computing, which trade higher power consumption for higher performance. In this paper we propose $\pi$-BA, the first hardware-software co-designed BA engine on an embedded FPGA-SoC that exploits custom hardware for higher performance and power efficiency. Specifically, based on our key observation that not all points appear on all images in a BA problem, we designed and implemented a Co-Observation Optimization technique to accelerate BA operations with optimized usage of memory and computation resources. Experimental results confirm that $\pi$-BA outperforms the existing software implementations in terms of performance and power consumption.
\end{abstract}


\IEEEpeerreviewmaketitle

\section{Introduction}

Bundle adjustment (BA) is the problem of refining a visual reconstruction to produce jointly optimal 3D structure and viewing parameter, including camera pose and calibration, estimates. Optimal means that the parameter estimates are found by minimizing some cost function that quantifies the model fitting error, and jointly means that the solution is simultaneously optimal with respect to both structure and camera variations \cite{triggs1999bundle}\cite{agarwal2010bundle}. Given a set of measured image feature locations and correspondences, the goal of BA is to find 3D point positions and camera parameters that minimize the reprojection error. This optimization problem is usually formulated as a non-linear least squares problem, where the error is the squared L2 norm of the difference between the observed feature location and the projection of the corresponding 3D point on the image plane of the camera. However, we are not limited to using the L2 norm; even when robust loss functions like Huber norm are used, the problem can be cast as a re-weighted non-linear least squares problem. In essence, BA is a large sparse geometric parameter estimation problem, the parameters being the combined 3D feature coordinates, camera poses and calibrations.

BA is widely used in many modern applications. Firstly, BA is the core component of 3D scene reconstruction applications: Agarwal \textit{et al.} present a system that can match and reconstruct 3D scenes from extremely large collections of photographs such as those found by searching for a given city on Internet photo sharing sites \cite{agarwal2009building}. The authors designed and implemented a cluster with 500 compute nodes to reconstruct cities consisting of 150K images in less than a day. In addition, BA is crucial in robotic localization applications:  Mur-Artal \textit{et al.} developed a feature-based simultaneous localization and mapping (SLAM) system, ORB-SLAM. The system consists of four modules, including tracking, mapping, relocalization, and loop closing. BA is used in the mapping stage for optimizing the visual feature map such that the robot can better localize itself \cite{mur2015orb}. Moreover, BA is used heavily in autonomous driving applications, especially in the production of high-definition maps \cite{liu2017creating}. BA is also used in space exploration mission as well, in multiple Mars exploration missions, NASA utilized BA technology to generate and optimize Mars explorer localization accuracies \cite{maimone2007two}. BA is also used in commercial products, such as Google street map, to perform scene reconstruction optimization \cite{klingner2013street}.

In both online real-time localization applications and offline visual reconstructions applications, BA remains the primary performance and power  consumption bottlenecks: for real-time localization systems (including mobile robots, autonomous vehicles, and space explorers) that perform local BA involving tens to hundreds of images, the latency of BA can be extremely high and thus fails to provide optimal localization updates in real-time. For offline visual reconstruction systems (including 3D scene reconstruction, street view maps, high-definition maps) that perform global BA involving thousands to millions of images, the power consumption of BA can be extremely costly. Previous approaches of optimizing BA performance heavily rely on parallel processing or distributed computing, which trade higher power consumption for higher performance. Nonetheless, to enable effective and efficient both online and offline applications, we need a BA solution that simultaneously optimize for performance and power  consumption, and thus we explore hardware acceleration techniques.

In this paper, aiming to achieve optimal performance and power efficiency for BA, we present $\pi$-BA, the first hardware-software co-designed BA engine on an embedded FPGA-SoC. The contribution of this paper is three-fold: first, this paper is the first exploration study of implementing a BA hardware accelerator, and the proposed $\pi$-BA's implementation has been proven effective. Second, based on our key observation that not all points appear on all images in a BA problem, we developed a novel Co-Observation Optimization technique for designing BA hardware accelerators. Third, in addition to achieving performance and power  efficiency, we also demonstrate that the proposed $\pi$-BA optimizes computing and memory resource usage.

The rest of this paper is organized as follows. In Section II, the related research works are presented. In Section III, we review the fundamentals of BA problems to help readers understand the challenges and complexities of designing BA hardware accelerators. In Sections IV and V we describe the $\pi$-BA architecture and delve into the novel Co-Observation Optimization design. In Section VI, we share the detailed experimental methodologies and results to demonstrate the effectiveness of $\pi$-BA architecture. Finally, we summarize the conclusions in Section VII.

\section{Related work}
In this section, we review several existing approaches of optimizing BA performance.  First, to optimize BA performance on CPU, Jeong \textit{et al.} exploit the block-sparsity pattern that arises in a reduced camera system and enhance the computational speed of the bundler with BLAS library matrix operations accelerations, efficient memory handling, and fast block-based linear solving \cite{jeong2012pushing}. Furthermore, the authors proposed novel embedded point iterations, which substantially improved the convergence speed by yielding a high cost decrease from each camera update step. In addition, the experimental results show the improved performance of the proposed bundler and provide useful and detailed comparisons among various choices when compositing a bundler.

Parallel processing using multicore, either on CPU or GPU, can be applied to optimize BA performance. Wu \textit{et al.} presented multicore solutions to the problem of bundle adjustment that run on currently available CPUs and GPUs \cite{wu2011multicore}. The authors concluded that using multicore systems deliver a 10x to 30x boost in speed over existing systems while reducing the amount of memory used. This was achieved by carefully restructuring the matrix vector product used in the PCG iterations into easily parallelizable operations. This restructuring also opens the door to a matrix free implementation which leads to substantial reductions in the memory consumption as well as execution time. The authors also showed that single precision arithmetic when combined with appropriate normalization gives numerical performance comparable to double precision based solvers while further reducing the memory and time cost. The resulting system enabled running the largest bundle adjustment problems to date on a single GPU.

Distributed computing is another effective way to optimize BA performance. Eriksson \textit{et al.} proposed a consensus framework to deal with large scale bundle adjustment in distributed system \cite{eriksson2016consensus}. Instead of merging small problems by the optimization of overlapping regions of small problems, the consensus framework utilizes the proximal splitting method to formulate the bundle adjustment problem, in which the small problems are merged by averaging points in fact, decreasing the cost of merging. The merging process for the same parameters guarantees the consensus of points in different nodes. This design may suffer from several problems. Firstly, in each iteration, each node in the distributed system has to broadcast all overlapping points to the master node to complete the merging process, which is a huge overhead for large scale data-sets. Secondly, parameters of each camera are independent of parameters of other cameras. However, in practice, some cameras may share the same intrinsic parameters. Thirdly, the method by merging points converges a little slowly in very large scale data-sets and may converge in a local minimum early.

Similarly, Zhang \textit{et al.} proposed a distributed approach for very large scale global bundle adjustment computation \cite{zhang2017distributed}. The proposed distributed formulation was derived from the classical optimization algorithm alternating direction method of multipliers, based on the global camera consensus. The authors analyzed the conditions under which the convergence of this distributed optimization would be guaranteed and they adopted over-relaxation and self-adaption schemes to improve the convergence rate. Also, the authors proposed to split the large scale camera-point visibility graph in order to reduce the communication overheads of the distributed computing.

The presented paper proposes $\pi$-BA, the first to BA hardware accelerator and its implementation on FPGA.  Compared to existing acceleration techniques, $\pi$-BA simultaneously optimize both performance and power  consumption, thus enabling both real-time local robotic localization applications and efficient offline visual reconstruction applications.

\section{Problem statement}
In the following sections, we use boldface to represent vectors and matrices.

\subsection{Perspective camera model}
\begin{figure}[t]
\centerline{\includegraphics[width=0.9\linewidth]{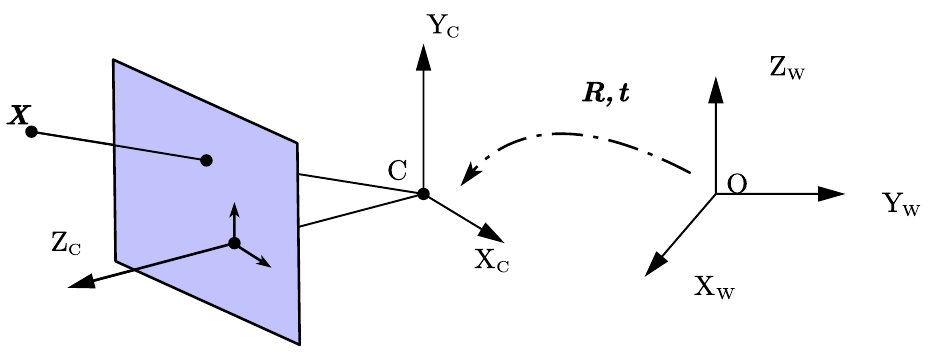}}
\caption{Perspective camera model. C and O are the origins of the camera coordinate and the world coordinate respectively.}
\label{Camera}
\end{figure}

In computer vision area, camera is a device that performs central projection of mapping 3D points onto a 2D image plane. Fig. \ref{Camera} illustrates the perspective camera model that projects a 3D points on a image plane. By employing projective geometry and coordinate transformation, the perspective projection is modeled by the following equation,

\begin{equation}
\boldsymbol{x}=s\boldsymbol{K}[\boldsymbol{R}|\boldsymbol{t}]\boldsymbol{X}
\label{perspective}
\end{equation}
where $\boldsymbol{X}$ is a $4\times1$ vector representing the position of a 3D points in the world coordinate, and $\boldsymbol{x}$ is a $3\times1$ vector representing the projection point's location in the image plane. Note that $\boldsymbol{X}$ and $\boldsymbol{x}$ are represented in the homogeneous coordinate. $[\boldsymbol{R}|\boldsymbol{t}]$ is a $3\times4$ matrix composed by a $3\times3$ rotation matrix $\boldsymbol{R}$ and a $3\times1$ translation vector $\boldsymbol{t}$. $\boldsymbol{R}$ and $\boldsymbol{t}$ are referred as extrinsic parameters of cameras and specify the rigid transformation from the world coordinate to the camera coordinate, as illustrated by Fig. \ref{Camera}. $\boldsymbol{K}$ is a $3\times3$ matrix of intrinsic parameters of cameras, which contains parameters such as focal length, focal position and etc. $s$ is a scale parameter.

\subsection{Bundle adjustment}
For visual simultaneous localization and mapping problems, bundle adjustment is employed in the last stage of the processing pipeline to further refine camera trajectories and 3D structures. It aims to minimize the discrepancies between observations of 3D points and predicted projections of the corresponding 3D points. Assume that $a$ 3D points are observed in $b$ images. Let $\boldsymbol{p_i}$ be the $i$-th 3D points, $\boldsymbol{o_{ij}}$ be the observation of the $i$-th point on the $j$-th image, and $\boldsymbol{c_j}$ be the $j$-th camera's parameter. $P(\boldsymbol{p_i}, \boldsymbol{c_j})$ denotes the projection function. Generally, bundle adjustment is formulated as a optimization problem, which is defined by Eq. \eqref{costfunction}. In the equation, $\sigma_{ij}$ evaluates to $1$ if the $i$-th 3D point is observed by the $j$-th camera, otherwise its value is $0$. This formulation shows that solving the bundle adjustment problem is to determine camera parameters and 3D points' positions such that observations are closely approximated by the corresponding re-projection points. Note that in the visual simultaneous localization and mapping problems, intrinsic parameters of cameras are known beforehand. As a result, only extrinsic parameters need to be optimized by bundle adjustment.

\begin{equation}
\underset{\boldsymbol{p_{i}},\boldsymbol{c_{j}}}{\mathop{\min}}\sum_{i=1}^{a}\sum_{j=1}^{b}{\sigma_{ij}{\parallel \boldsymbol{o_{ij}}-P\left(\boldsymbol{p_i},\boldsymbol{c_j}\right)\parallel}}
\label{costfunction}
\end{equation}

\begin{algorithm}[t]
\scriptsize
\caption{Levenberg-Marquardt Method}
\label{alg: LM}
\begin{algorithmic}
\STATE {\bf Input:} A vector function $f:\mathbb{R}^{m}\rightarrow\mathbb{R}^{n}$ with $n>m$, \\
~~~~~~~~~a measurement vector $\mathbf{x}\in\mathbb{R}^{m}$, \\
~~~~~~~~~initial parameter $\mathbf{p_0}\in\mathbb{R}^{n}$.
\STATE {\bf Output:} A vector $\mathbf{p^{+}}\in\mathbb{R}^{m}$ minimizing ${\|\mathbf{x}-f(\mathbf{p})\|}^2$.
\STATE {\bf Algorithm:} \\
\STATE {1:}~$k:=0;v:=2;\mathbf{p}:=\mathbf{p_0};$ \\
\STATE {2:}~$\mathbf{J}:=f'(\mathbf{p});\epsilon:=\mathbf{x}-f(\mathbf{p});$
\STATE {3:}~$\mathbf{A}:=\mathbf{J^{T}J};\mathbf{D^TD}=\mbox{diag}\mathbf{(A)};\mathbf{g}:=\mathbf{J^{T}\epsilon};$\\
\STATE {4:}~$\mbox{stop}:=(\|\mathbf{g}\|_\infty\leq\varepsilon_1);\mu:=\tau*\mbox{max}_{i=1,\cdots,m}(A_{ii})$\\
\STATE {5:}~$\mbox{while }(\mbox{not stop}) \mbox{ and } (k<k_{max})$\\
\STATE {6:}~~~~~$k:=k+1;$\\
\STATE {7:}~~~~~$\boxed{\mbox{Solve}(\mathbf{A}+\mu\mathbf{D^TD})\delta_{\mathbf{p}}=\mathbf{g};}$\\
\STATE {8:}~~~~~$\mbox{if}(\|\delta_{\mathbf{p}}\|\leq\varepsilon_2\|\mathbf{p}\|)$\\
\STATE {9:}~~~~~~~~~$\mbox{stop}:=\mbox{true};$\\
\STATE {10:}~~~~$\mbox{else}$\\
\STATE {11:}~~~~~~~~$\mathbf{p}_{new}:=\mathbf{p}+\delta_{\mathbf{p}};$\\
\STATE {12:}~~~~~~~~$\rho:=(\|\epsilon\|^{2}-\|\mathbf{x}-f(\mathbf{p_{new}})\|^{2})/(\delta_{\mathbf{p}}^{T}(\mu\delta_{\mathbf{p}}+\mathbf{g}));$\\
\STATE {13:}~~~~~~~~$\mbox{if}(\rho>0)$\\
\STATE {14:}~~~~~~~~~~~~$\mathbf{p}:=\mathbf{p}_{new};$\\
\STATE {15:}~~~~~~~~~~~~$\mathbf{J}:=f'(\mathbf{p});\epsilon_:=\mathbf{x}-f(\mathbf{p});$\\
\STATE {16:}~~~~~~~~~~~~$\mathbf{A}:=\mathbf{J^{T}J};\mathbf{D^TD}=\mbox{diag}\mathbf{(A)};\mathbf{g}:=\mathbf{J^{T}\epsilon};$\\
\STATE {17:}~~~~~~~~~~~~$\mbox{stop}:=(\|\mathbf{g}\|_\infty\leq\varepsilon_1);$\\
\STATE {18:}~~~~~~~~~~~~$\mu:=\mu*\mbox{max}(\frac{1}{3},1-(2\rho-1)^{3});v:=2;$\\
\STATE {19:}~~~~~~~~$\mbox{else}$\\
\STATE {20:}~~~~~~~~~~~~$\mu:=\mu*v;v:=2*v;$\\
\STATE {21:}~~~~~~~~$\mbox{endif}$\\
\STATE {22:}~~~~$\mbox{endif}$\\
\STATE {23:}$\mbox{endwhile}$\\
\STATE {24:}$\mathbf{p}^{+}:=\mathbf{p}$\\
\end{algorithmic}
\end{algorithm}

\subsection{Levenberg-Marquardt's algorithm}
Levenberg-Marquardt's (LM) algorithm is a non-linear least squares algorithm. It is widely used to find a local minimum of the functions that are expressed as a sum of squares of several nonlinear functions. LM combines the merits of the steepest descent and the Gauss-Newton method. It can converge from a wide range of initial conditions. LM has become a standard algorithm for performing the bundle adjustment in visual SLAM and 3D reconstruction problems \cite{triggs1999bundle}\cite{strasdat2010real}.

Algorithm \ref{alg: LM} shows the pseudo code of LM algorithm. In the pseudo code, ${\parallel\bullet\parallel}$ and ${\parallel\bullet\parallel_\infty}$ denote the 2 and infinity norm, respectively. $\bullet^T$ is matrix transposition operator. Assume that there are $a$ 3D points, $b$ cameras and $o$ observations. The inputs of the algorithm are a $n\times1$ measurement vector $\mathbf{x}$ and a $m\times1$ initial parameter vector $\mathbf{p_0}$. According to Eq.~\eqref{Camera}, it can be derived that $m$ equals $3a+6b$ and $n$ equals $2o$. $f$ is a vector function that maps a parameter vector $\mathbf{p}$ to an estimated measurement vector. The output of the algorithm is an optimized parameter vector $\mathbf{p^+}$ that minimize $\mathbf{\epsilon}^T\mathbf{\epsilon}$, where $\mathbf{\epsilon}=\mathbf{x} - f(\mathbf{p})$. $\mathbf{J}$ is the Jacobian matrix of $f(\mathbf{p})$.

LM algorithm solves a nonlinear optimization problem by iteratively linearizing the nonlinear function and solving the linearized equation. In each iteration, it firstly computes the change of $\mathbf{p}$, namely $\mathbf{\delta_\mathbf{p}}$, through solving the linear equation, and then updates $\mathbf{p}$. The stop conditions of LM algorithm are: 1) the magnitude of gradient, $\|\mathbf{g}\|_\infty$, is less than $\varepsilon_1$; 2) the change of magnitude of $\mathbf{p}$, $\|\delta_{\mathbf{p}}\|$, is less than $\varepsilon_2\|\mathbf{p}\|$; 3) the maximum iteration step, $k_{max}$, is reached. $\varepsilon_1$, $\varepsilon_2$, $k_{max}$ and $\tau$ are parameters specified by users. More details of LM algorithm for bundle adjustment can be found in \cite{Lourakis2005Is}.

In the LM algorithm, the Jacobian matrix, $\mathbf{J}$, is a $(3a+6b)\times 2o$ matrix. $\mathbf{J^TJ}+\mu\mathbf{D^TD}$ is a $(3a+6b)\times(3a+6b)$ square matrix. Directly solving Eq. \eqref{eq:normalEquation} requires $(3a+6b)^3$ arithmetic operations, which is computationally intensive. In practice, matrix elimination technique and Cholesky factorization are used to reduce the computational complexity of solving Eq. \eqref{eq:normalEquation}.

\begin{equation}\label{eq:normalEquation}
\left(\mathbf{A}+\mu\mathbf{D^TD}\right)\delta\mathbf{p}=\mathbf{g}
\end{equation}

The parameter vector $\mathbf{p}$ can be divided into a 3D points part and a camera parameter part, and is expressed as $\mathbf{p}=[\mathbf{p}^p; \mathbf{p}^c]$. Similarly, the Jacobian matrix $\mathbf{J}$ can be divided into a 3D points Jacobian matrix and a camera parameter Jacobian matrix, as shown by the following equation. $\mathbf{J}^{p}$ and $\mathbf{J}^{c}$ represent the Jacobians of 3D points and cameras, respectively.

\begin{equation}
\mathbf{J}=
\begin{bmatrix}
\begin{array}{c|c}
\begin{matrix}
\mathbf{J}^p
\end{matrix}
&
\begin{matrix}
\mathbf{J}^c
\end{matrix}
\end{array}
\end{bmatrix}
\label{twopart}
\end{equation}

Given that $\mathbf{A}=\mathbf{J^{T}J}$ and $\mu\mathbf{D^TD}=\mbox{diag}\mathbf{(A)}$. By combining Eq. \eqref{twopart}, $\mathbf{A}+\mu\mathbf{D^{T}D}$ can be expressed by a simple block matrix, shown in the following equation. $\mathbf{U}$ and $\mathbf{V}$ are a $3a\times 3a$ and a $6b\times 6b$ matrix. $\mathbf{W}$ is a $6b\times 3a$ matrix.

\begin{algorithm}[t]
\scriptsize
\caption{Schur Elimination}
\label{alg:Schur}
\begin{algorithmic}
\STATE {\bf Input:} Jacbian matrix of reprojection error function $\mathbf{J}$,\\
~~~~~~~~~residual vector, reprojection error $\mathbf{\epsilon}$, \\
~~~~~~~~~trust region radius $\sqrt{\mu}\mathbf{D}$, \\
\STATE {\bf Output:} matrix $\mathbf{S}$,\\
~~~~~~~~~~~vector $\mathbf{r}$
\STATE {\bf Algorithm:} \\
\STATE {1:}~for $j=1$ to $b$\\
\STATE {2:}~~~~~$\mathbf{S}_{jj}:=\mu\mathbf{D}_{j}^{c\mathbf{T}}\mathbf{D}_{j}^{c}$\\
\STATE {3:}~for $i=1$ to $a$\\
\STATE {4:}~~~~~$\mathbf{U}_{i}:=\mu\mathbf{D}_{i}^{p\mathbf{T}}\mathbf{D}_{i}^{c}$\\
\STATE {5:}~~~~~$\mathbf{g}_{i}^{p}:=\mathbf{0}$\\
\STATE {6:}~~~~~for $j=1$ to $b$\\
\STATE {7:}~~~~~~~~~$\mathbf{U}_{i}:=\mathbf{U}_{i}+\mathbf{J}_{ij}^{p\mathbf{T}}\mathbf{J}_{ij}^{p}$\\
\STATE {8:}~~~~~~~~~$\mathbf{g}_{i}^{p}:=\mathbf{g}_{i}^{p}+\mathbf{J}_{ij}^{p\mathbf{T}}\mathbf{\epsilon}_{ij}$\\
\STATE {9:}~~~~~~~~~$\mathbf{W}_{ij}:=\mathbf{J}_{ij}^{c\mathbf{T}}\mathbf{J}_{ij}^{p}$
\STATE {10:}~~~~~~~~$\mathbf{S}_{jj}:=\mathbf{S}_{jj}+\mathbf{J}_{ij}^{c\mathbf{T}}\mathbf{J}_{ij}^{c}$\\
\STATE {11:}~~~~~~~~$\mathbf{r}_{j}:=\mathbf{r}_{j}+\mathbf{J}_{ij}^{c\mathbf{T}}\mathbf{\epsilon}_{ij}$
\STATE {12:}~~~~$\mathbf{inv}:=\mathbf{U}_{i}^{\mathbf{-1}}$\\
\STATE {13:}~~~~for $j_{1}=1$ to $b$\\
\STATE {14:}~~~~~~~~$\mathbf{r}_{j_1}:=\mathbf{r}_{j_1}-\mathbf{W}_{ij_1}\times\mathbf{inv}\times\mathbf{g}_{i}^{p}$\\
\STATE {15:}~~~~~~~~for $j_{2}=1$ to $b$\\
\STATE {16:}~~~~~~~~~~~~$\mathbf{S}_{j_1j_2}:=\mathbf{S}_{j_1j_2}-\mathbf{W}_{ij_1}\times\mathbf{inv}\times\mathbf{W}_{ij_2}^{\mathbf{T}}$\\
\end{algorithmic}
\end{algorithm}

\begin{multline}
\mathbf{A}+\mu\mathbf{D^TD}
=
\begin{bmatrix}
\mathbf{J}^{p\mathbf{T}}\mathbf{J}^p & \mathbf{J}^{p\mathbf{T}}\mathbf{J}^c\\
\mathbf{J}^{c\mathbf{T}}\mathbf{J}^p & \mathbf{J}^{c\mathbf{T}}\mathbf{J}^c
\end{bmatrix}
+\mu
\begin{bmatrix}
\mathbf{D}^{p\mathbf{T}}\mathbf{D}^p & 0                                  \\
0                                    & \mathbf{D}^{c\mathbf{T}}\mathbf{J}^c
\end{bmatrix} \\
=
\begin{bmatrix}
\mathbf{U} & \mathbf{W^T} \\
\mathbf{W} & \mathbf{V}
\end{bmatrix}
\label{JTJ_Structure}
\end{multline}

By combining Eq.~\eqref{eq:normalEquation} and Eq.~\eqref{JTJ_Structure}, we obtain the following equation.

\begin{equation}
\begin{bmatrix}
\mathbf{U} & \mathbf{W^{T}}\\
\mathbf{W} & \mathbf{V}
\end{bmatrix}
\begin{bmatrix}
\mathbf{\delta p}^{p}\\
\mathbf{\delta p}^{c}
\end{bmatrix}
=
\begin{bmatrix}
\mathbf{J}^p\\
\mathbf{J}^c
\end{bmatrix}
\epsilon
\label{simply}
\end{equation}

By eliminating the lower left block, $\mathbf{W}$, the following equation is obtained. In the equation, $\mathbf{V-WU^{-1}W^T}$ is Schur complement matrix, which is a symmetric matrix and is denoted as $\mathbf{S}$ in this paper. Vector $(\mathbf{J}^c-\mathbf{WU^{-1}J}^p)\epsilon$ is denoted as $\mathbf{r}$ in this paper. Then, $\delta\mathbf{p}^c$ can be obtained by solving Eq.~\eqref{eq:solve}. In practice, $\delta\mathbf{p}^c$ is solved by Cholesky factorization.

\begin{equation}
(\mathbf{V-WU^{-1}W^T})\delta\mathbf{p}^c=(\mathbf{J}^c-\mathbf{WU^{-1}J}^p)\epsilon
\label{eq:solve}
\end{equation}

After obtaining $\delta\mathbf{p}^c$, the change of 3D points vector, $\delta\mathbf{p}^p$, can be obtained by back substitution. Eq. \eqref{delta_p} describes the closed-form solution of $\delta\mathbf{p}^p$. Note that $\mathbf{U}$ is a diagonal block matrix, of which diagonal elements are $3 \times 3$ matrices. The cost of computing the inversion of $\mathbf{U}$ is low.

\begin{equation}
\delta\mathbf{p}^p=\mathbf{U^{-1}}(\mathbf{J}^{p\mathbf{T}}\epsilon-\mathbf{W}\delta\mathbf{p}^c)
\label{delta_p}
\end{equation}

The computation of $\mathbf{S}$ is called Schur elimination in this paper. Directly calculating $\mathbf{S}$ according to its expression, $\mathbf{V-WU^{-1}W^T}$, is computationally expensive. Since $\mathbf{U}$, $\mathbf{V}$ and $\mathbf{W}$ are sparse matrices, the complexity of calculating $\mathbf{S}$ can be substantially reduced by exploiting the structure of these sparse matrices. Algorithm \ref{alg:Schur} describes the procedure of calculating $\mathbf{S}$ and $\mathbf{r}$.
According to the algorithm, computing $\mathbf{S}$ requires $ab^2$ arithmetic operations. Employing Cholesky factorization to solve Eq. \eqref{eq:solve} requires $(6b)^3/3$ operation. For vSLAM problems and 3D reconstruction problems, the number of 3D points $a$ is much larger than the number of images $b$. The Shur elimination is the most computationally intensive step when solving the linear equation in LM algorithm.

\section{System architecture of LM implementation}

\begin{figure}[t]
\centerline{\includegraphics[width=0.9\linewidth]{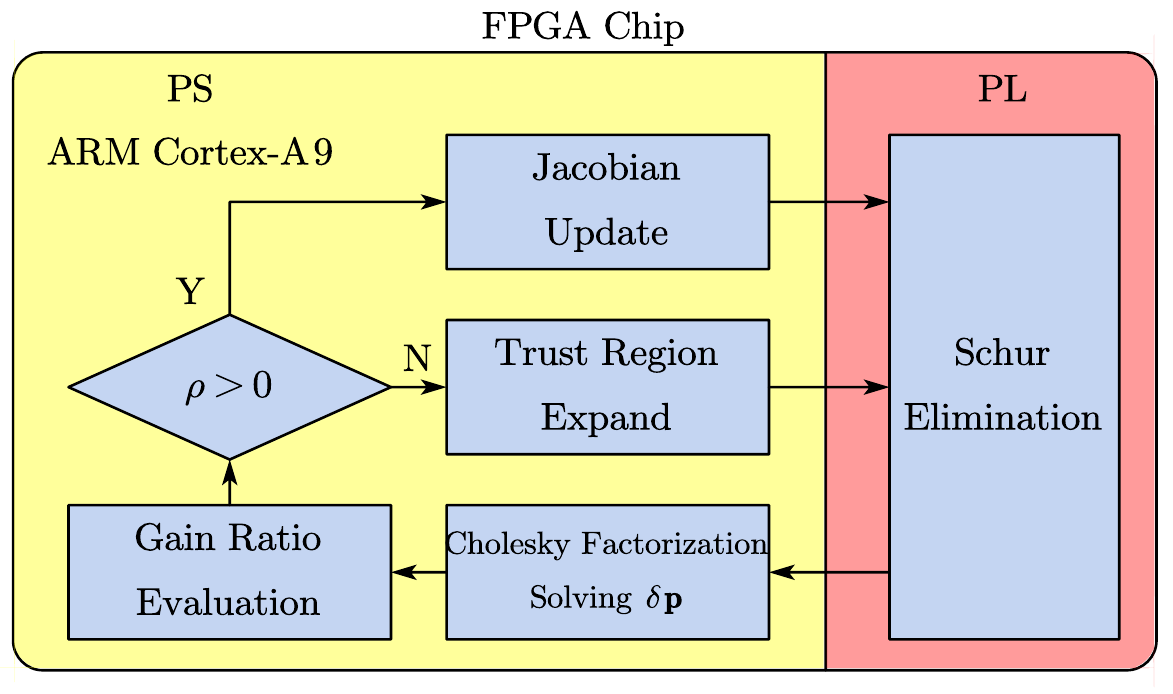}}
\caption{System architecture}
\label{co-design}
\end{figure}

After introducing Schur elimination, the computations in one iteration of LM algorithm can be divided into five parts including Jacobian update (JU), Schur elimination (SE), Cholesky factorization sloving $\delta\mathbf{p}$ (CFS), gain ratio evaluation (GRE) and trust region expand (TRE). One of the most time-consuming parts is SE, which has complexity $O(ab^2)$. Therefore, we propose a hardware-software co-design in which the SE is accelerated in hardware and other parts are implemented in software. The whole system architecture is shown in Fig. \ref{co-design}. The amount of data transferred between hardware and software is $18o+b^2/2$. We use AXI Direct Memory Access (DMA), 6400Mbit/s. The measured performance shows that the data transfer time is less than the hardware computation time, and also both data transfer and computation are pipelined to reduce the data transfer overhead.

\begin{figure*}[t]
\centering
\subfigure[Jacobian matrix]{
\begin{minipage}[t]{0.3\linewidth}
\centering
\includegraphics[width=1\linewidth]{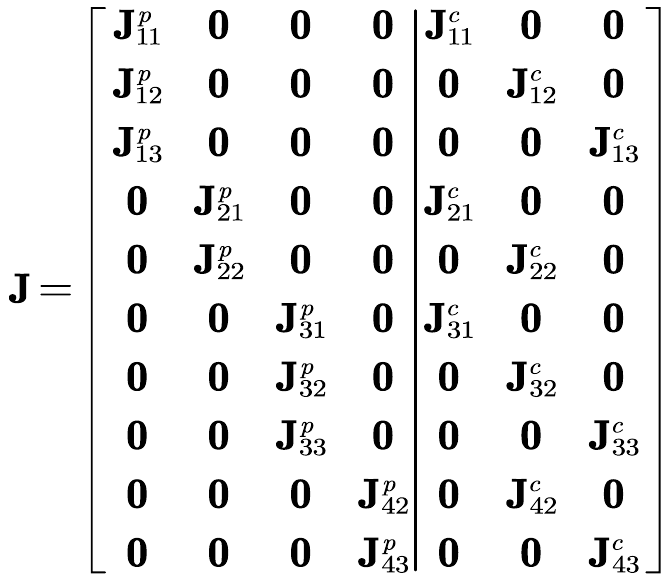}
\label{fig:JForm}
\end{minipage}
}
\subfigure[Storage format conversion of Jacobian matrix]{
\begin{minipage}[t]{0.5\linewidth}
\centering
\includegraphics[width=1\linewidth]{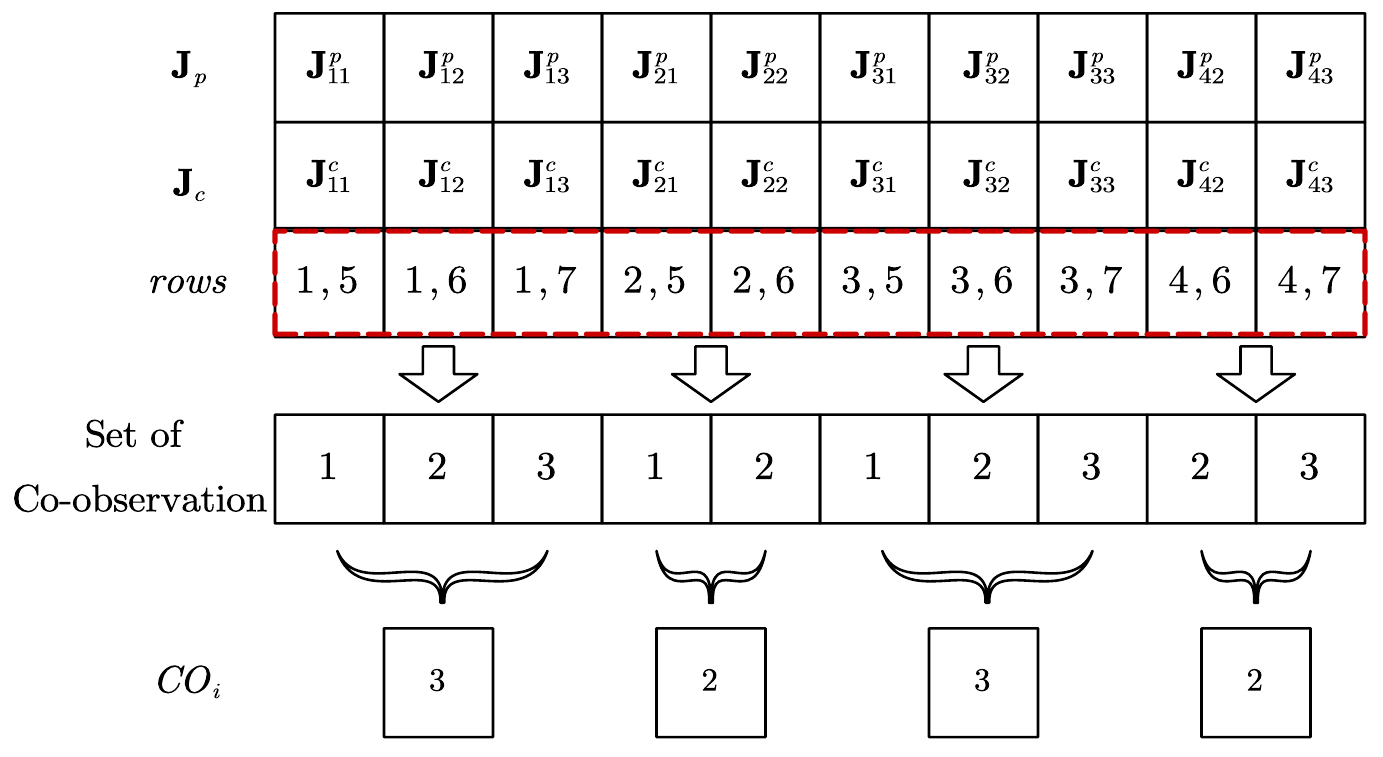}
\label{fig:JStorge}
\end{minipage}
}
\centering
\caption{Storage format conversion of Jacobian matrix}
\label{fig:Jacobian_store}
\end{figure*}

\begin{figure*}[t]
\centering
\subfigure[Diagonal of matrix $\mathbf{S}$]{
\begin{minipage}[t]{0.25\linewidth}
\centering
\includegraphics[width=0.80\linewidth]{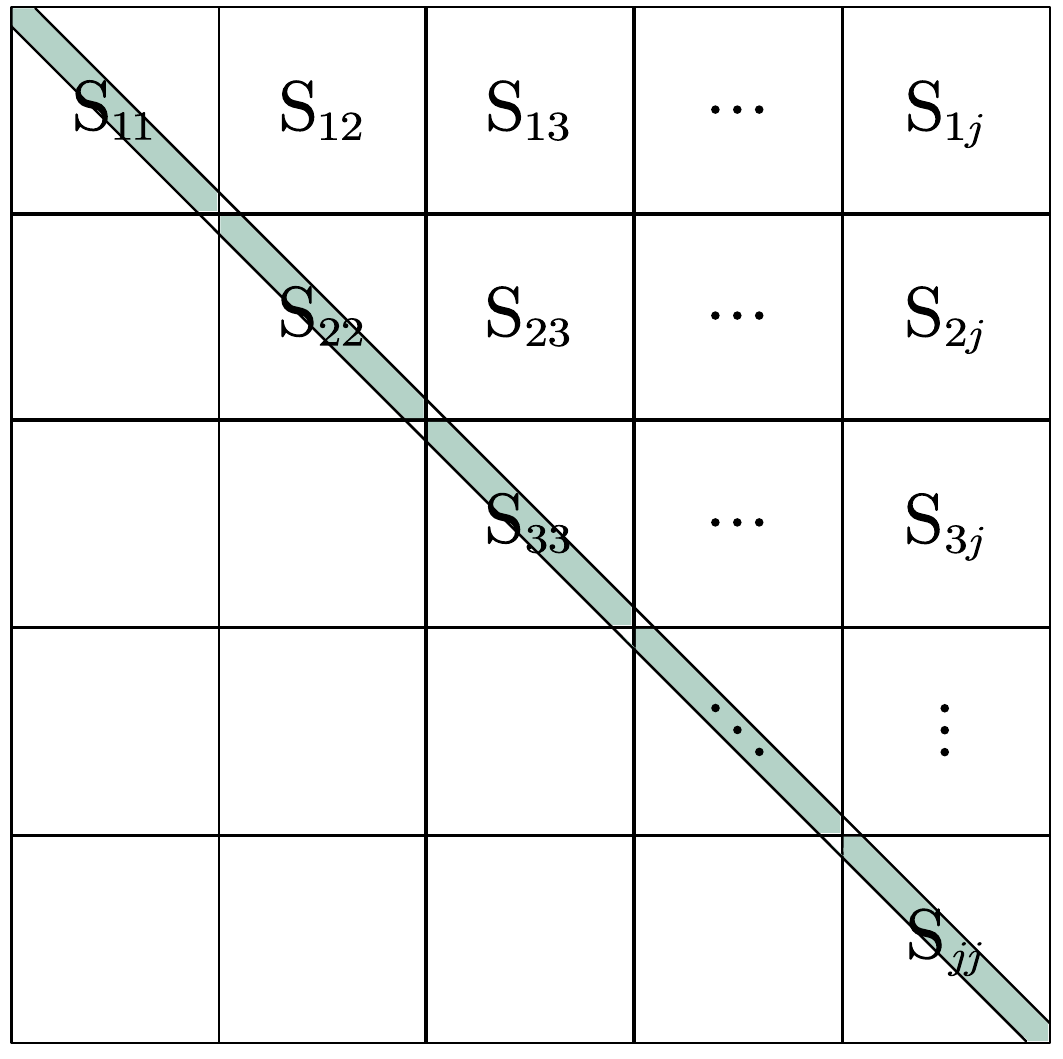}
\end{minipage}
}
\subfigure[Diagonal block of matrix $\mathbf{S}$]{
\begin{minipage}[t]{0.25\linewidth}
\centering
\includegraphics[width=0.80\linewidth]{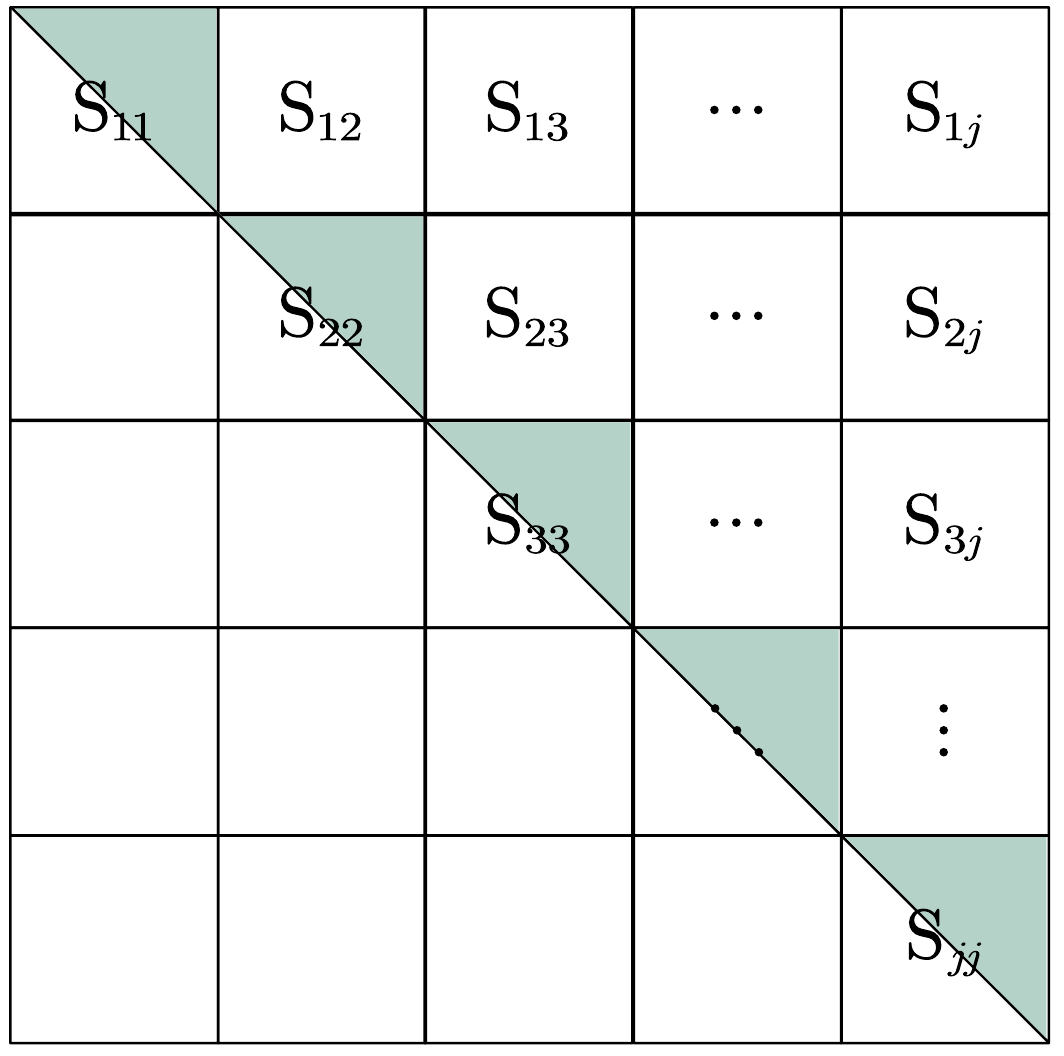}
\end{minipage}
}
\subfigure[Matrix $\mathbf{S}$]{
\begin{minipage}[t]{0.25\linewidth}
\centering
\includegraphics[width=0.80\linewidth]{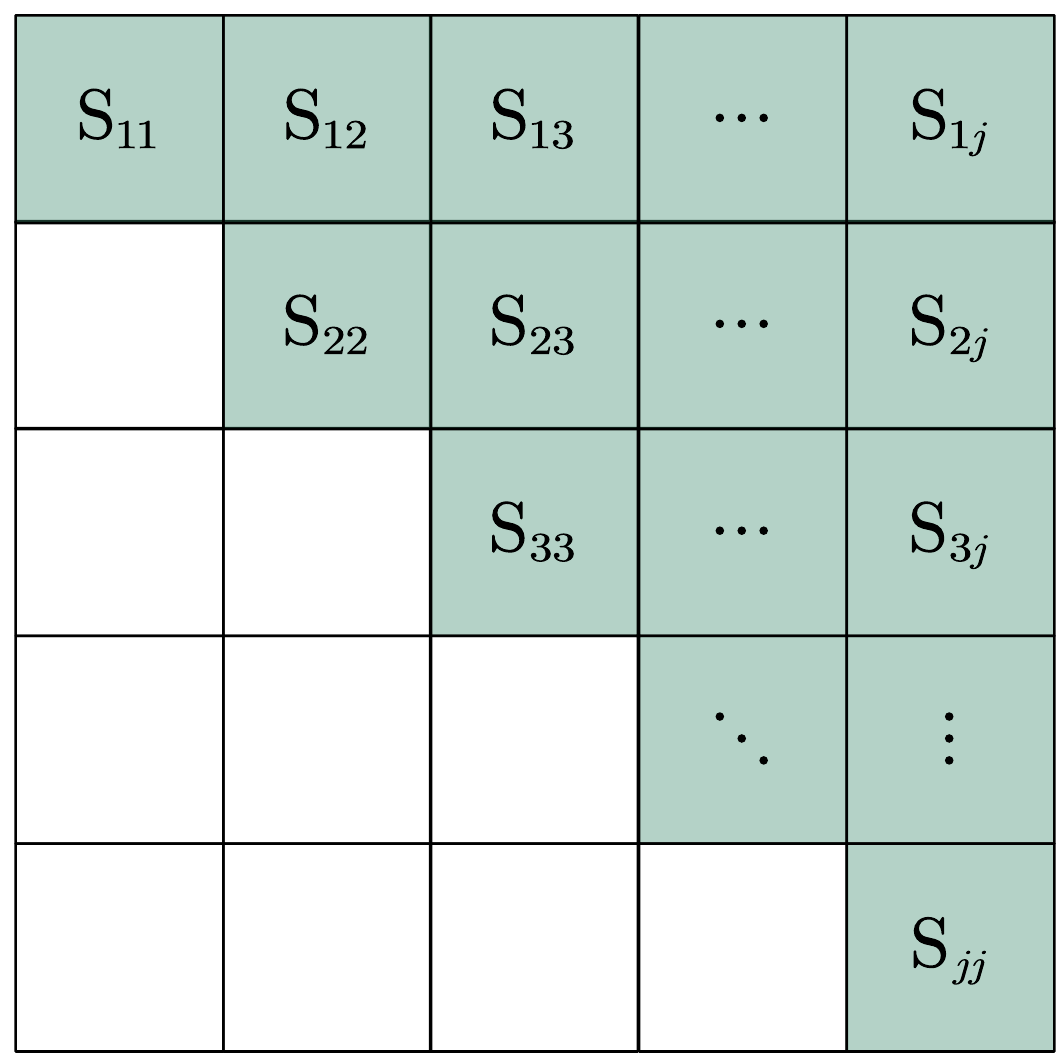}
\end{minipage}
}
\centering
\caption{Storage of matrix $\mathbf{S}$}
\label{sstorge}
\end{figure*}

\begin{figure*}[t]
\centerline{\includegraphics[width=0.9\linewidth]{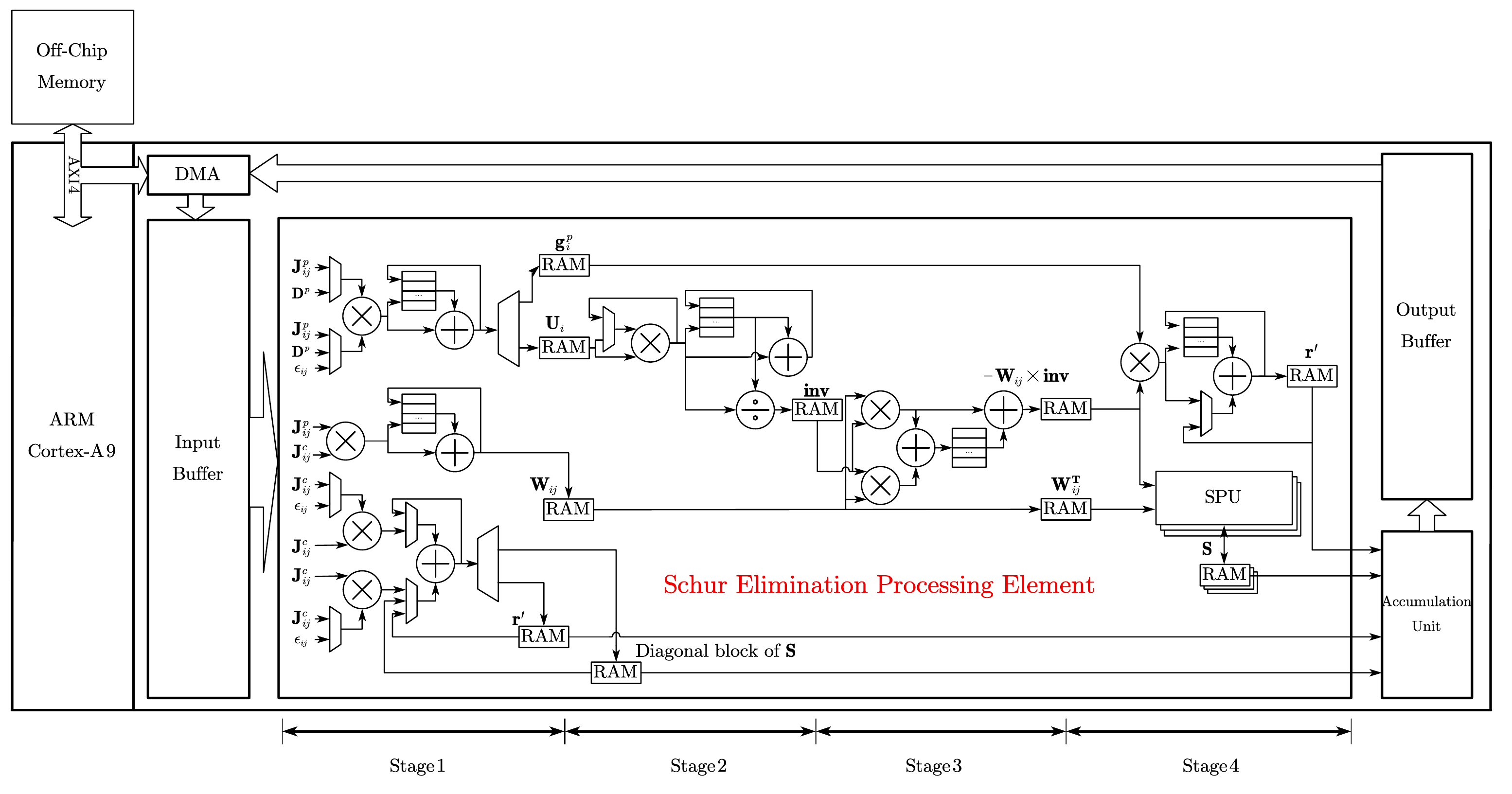}}
\caption{Hardware architecture of bundle adjustment}
\label{Architecture}
\end{figure*}

As shown in Algorithm \ref{alg:Schur}, the SE mainly computes $\mathbf{S}~(6b \times 6b)$ given input $\mathbf{J}~(2o \times (3a+6b))$ and $\mathbf{D}~(3a+6b)$. Given the large matrices, data storage format is a key to system performance and size.

The Jacobian matrix is a block sparse matrix, like Fig. \ref{fig:JForm}.
Therefore, Jacobian matrix uses a block compressed sparse row (BCSR) storage format \cite{barrett1994templates}. Typical BCSR format uses three attributes: \textit{values} storing linearly in row-wise order the values of all blocks, \textit{rows} storing the block-column indices in a row, and \textit{block-starts} storing the indices of the first element of each block in values. In this paper, because the block size of $\mathbf{J}$ is fixed in BA, the block-starts information is not needed. In addition, we convert the \textit{rows} attributes to a set of co-observations, which means a set of images observing a point $i$, and the number of elements in the set ($CO_i$) which is called co-observation value and indicates that point $i$ is observed by how many images. In Fig.~\ref{fig:JForm}, the set of co-observations of the second point is $\{1,2\}$ and $CO_2=2$ because the second is observed by the first and second images. The conversion is shown in Fig.~\ref{fig:JStorge}. The advantages of doing this are that 1) The set and $CO_i$ can address \textit{values}, and 2) they contain the physical meaning of the BA problem and  can be directly used in subsequent computations without the need of calculation on-the-fly in hardware. In addition, $\mathbf{J}^p$ and $\mathbf{J}^c$ are stored separately in off-chip memory.

The $\mathbf{S}$ matrix is a dense matrix and obtained by accumulating the sub-matrices continuously as shown in Algorithm \ref{alg:Schur}. In the entire calculation, the $\mathbf{S}$ matrix is divided into three parts. The first part is a diagonal matrix calculated in Line 2 of Algorithm 2, which is stored as a vector shown in Fig.~\ref{sstorge}(a). The second part is a diagonal block matrix computed in Line 9 of Algorithm 2. Only a half of the diagonal blocks is stored shown Fig.~\ref{sstorge}(b). The last part is a partial accumulation of $\mathbf{S}$ matrix as shown in Line 15 of Algorithm 2. Because $\mathbf{S}$ matrix is symmetric positive definite, nearly half of $\mathbf{S}$ is stored shown in Fig.~\ref{sstorge}(c). In order to simplify the hardware control design, the whole diagonal blocks are preserved to maintain computation regularity on the diagonal blocks and other blocks.

\section{FPGA Implementation of Schur Elimination with Co-observation optimization}

The customized SE hardware implementation is sketched in Fig. \ref{Architecture}. The input buffer temporarily stores the data of the Jacobian matrix transmitted from the off-chip memory. The middle is the main computation of SE, called SE processing element. The accumulation unit computes the Line 2 of Algorithm \ref{alg:Schur} and adds the intermediate computational results of $\mathbf{S}$  and $\mathbf{r}$, respectively. The output buffer temporarily stores $\mathbf{S}$ and $\mathbf{r}$  which need to be transferred back to the off-chip memory. In the next, we present the detailed design of the SE processing element (PE). Because the co-observation value $CO_i$ has impacts on the speed, computation and memory resource usage of the PE, we propose a Co-Observation Optimization design method.

As shown in Fig. \ref{Architecture}, the PE is partitioned into four stages according to data dependencies and as early as possible scheduling. The first stage performs the computations of Lines 4-11 in Algorithm \ref{alg:Schur}. In the second stage, the inverse of $\mathbf{U}_i$ is calculated, corresponding Line 12 of the algorithm. The matrix computation $-\mathbf{W}_{ij} \times \mathbf{inv}$ is completed in the third stage. The fourth stage completes the remainder computations of Lines 14 and 16. The matrix multiplications and additions in the matrix $\mathbf{S}$ processing unit (SPU) in the fourth stage are fully parallelized.

\begin{figure}[t]
\centerline{\includegraphics[width=0.7\linewidth]{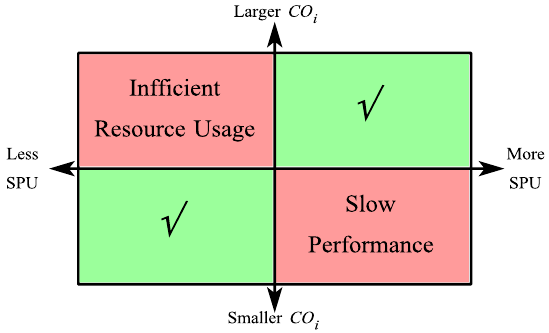}}
\caption{Impacts of match degree between the number of SPUs and $CO_i$ on efficiency}
\label{SPUandCOi}
\end{figure}

The latencies of the four stages are affected by $CO_i$. The latencies of the first and third stages are $36CO_i$ cycles. The latency of the fourth stage is $18(CO_i^2+CO_i)$ cycles. The latency of the second stage is 70 cycles, independent of $CO_i$. The fourth stage is the bottleneck. To speedup the operations, SPU is duplicated for parallel computation on different data. The number of duplications depends on $CO_i$ and available hardware resources. Given sufficient resources, the matching efficiency between the number of SPUs and $CO_i$ is shown in Fig.~\ref{SPUandCOi}. We can see that mismatch of both may lead to slow performance or inefficient resource usage. In practice, $CO_i$ varies significantly among points, which will be shown in the result section. Therefore, the structure of PE should be optimized to match to the majority of the $CO_i$ of points.

In our tested data, more than 30\% points have $CO_i=2$. We customize one PE for mainly processing points with $CO_i=2$. The number of SPUs ($q$) of the PE is roughly determined as
\begin{equation}
18(CO_i^2+CO_i)/q \approx 36CO_i \approx 70
\label{COiEq}
\end{equation}
resulting in $q=2$.

In addition, the amount of intermediate computational results is related to $CO_i$ as shown in Lines 7-11 of Algorithm \ref{alg:Schur}. As a result, the size of RAMs in a PE is determined by the maximum $CO_i$ processed by the PE. In the customized PE for $CO_i=2$ the on-chip memory usage can be reduced.

Ideally, PEs can be customized with different $q$ for the points with larger $CO_i$ accordingly. In practice, $q$ is determined not only by Eq. \eqref{COiEq}, but also by available resources. When the available resources are not enough, SPU duplication is not feasible. In this case, the first three stages of PE can be slowed down to save computational resources. For example, one multiplier and one adder can be removed from the third stage when the fourth stage is the bottleneck.

Moreover, the second stage involves matrix inverse computation. In the original software implementation, Cholesky factorization is used, containing complex operations such as square root and division. Especially, multiple square root operations have interdependencies and can not be parallelized. To solve the issue, we use determinant and adjoint matrix to invert the matrix $\mathbf{U}_{i}^{\mathbf{-1}}=\frac{\mbox{adj}\mathbf{U}_{i}}{\mbox{det}\mathbf{U}_{i}}$, given the small $3 \times 3$ matrix.

\begin{figure}[t]
\centerline{\includegraphics[width=0.8\linewidth]{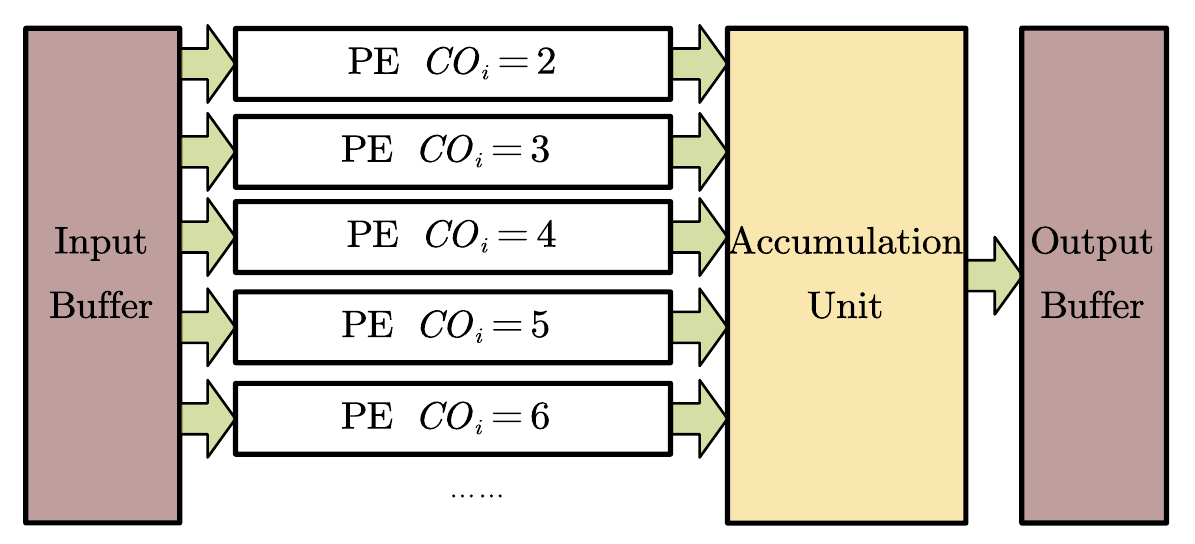}}
\caption{Co-observation optimization}
\label{co-ob-idea}
\end{figure}

\section{Experimental results}

The proposed hardware-software co-designed BA implementation is evaluated on a Zynq xc7z030sbg485-1 FPGA platform. The software part using double-precision floating-point numerical representation is executed on an ARM Cortex-A9@667Hz. The hardware part using single-precision floating-point numerical representation is synthesized and implemented on the programmable logic with maximum clock frequency 180 MHz. For the hardware part, three versions are implemented: one PE with one SPU (named Schur\_1), one PE with two SPUs (named Schur\_2), and two PEs each having two SPUs (named Schur\_3).

Our implementation is compared with a software implementation which uses the non-linear optimization library Ceres-Solver \cite{ceres-solver}. The software implementation using double-precision floating-point numerical representation runs on a Intel Pentium G2030 CPU at 3.0 GHz with 4 GB of RAM and on ARM Cortex-A9 core, respectively.

The datasets used in our experiments are from Bundle Adjustment in the Large (BAL) \cite{agarwal2010bundle}. We choose five datasets, each containing input images below 50. This is because local BA on 50 images is enough for SLAM application and for simple scene SfM. The number of images, points and observations of the five datasets are shown in Table~\ref{table:data}. The five datasets contain different combinations of images, points and observations, and represent different scenes.

\begin{table}[t]
\caption{Datasets used for experiments}
\begin{center}
\begin{tabular}{|c|c|c|c|}
\hline
{\textbf{Dataset}} & {\textbf{Images}} & {\textbf{Points}} & {\textbf{Observations}} \\
\hline
\textcircled{1} & 16  & 22106 &  83718 \\
\hline
\textcircled{2} & 21  & 11315 &  36455 \\
\hline
\textcircled{3} & 39  & 18060 &  63551 \\
\hline
\textcircled{4} & 49  & 7776  &  31843 \\
\hline
\textcircled{5} & 50  & 20431 &  73967 \\
\hline
\end{tabular}
\label{table:data}
\end{center}
\end{table}

\begin{table*}[t]
\caption{Distribution of co-observation values $CO_i$ of the datasets}
\begin{center}
\begin{tabular}{|c|c|c|c|c|c|c|c|c|c|c|c|c|c|c|c|c|c|c|c|c|c|}
\hline
{} & \multicolumn{10}{|c|}{\textbf{Percentage of specific co-observation values(\%)}}\\
\hline
$CO_i$ &2     &3     &4     &5     &6     &7     &8     &9     &10    &11      \\
\hline
\textcircled{1}    &38.89 &20.41 &12.96 &9.12  &6.39  &4.32  &2.92  &2.04  &1.38  &0.94    \\
\hline
\textcircled{2}    &51.78 &19.60 &10.97 &6.57  &4.05  &2.85  &1.74  &1.09  &0.65  &0.27    \\
\hline
\textcircled{3}    &51.86 &17.19 &10.03 &5.69  &4.06  &3.04  &2.31  &1.81  &1.21  &0.96    \\
\hline
\textcircled{4}    &44.35 &17.84 &10.60 &6.73  &5.00  &3.33  &2.73  &2.13  &1.62  &1.53    \\
\hline
\textcircled{5}    &52.14 &17.08 &9.30  &5.51  &3.93  &2.69  &2.40  &1.75  &1.48  &1.02    \\
\hline\hline
$CO_i$ &12    &13    &14    &15    &16    &17    &18    &19    &20    &21$\sim$29\\
\hline
\textcircled{1}    &0.44  &0.17  &0.03  &0     &0     &0     &0     &0     &0     &0     \\
\hline
\textcircled{2}    &0.18  &0.15  &0.07  &0.04  &0     &0     &0     &0     &0     &0     \\
\hline
\textcircled{3}    &0.64  &0.35  &0.42  &0.22  &0.13  &0.06  &0.01  &0.01  &0.01  &0     \\
\hline
\textcircled{4}    &1.02  &0.64  &0.51  &0.35  &0.40  &0.35  &0.17  &0.21  &0.12  &0.39  \\
\hline
\textcircled{5}    &0.65  &0.57  &0.48  &0.35  &0.29  &0.12  &0.09  &0.09  &0.02  &0.02  \\
\hline
\end{tabular}
\label{co-ob-table}
\end{center}
\end{table*}

As mentioned earlier, co-observation value $CO_i$ affects the speed, efficiency and RAM usage of the Schur elimination processing element. Therefore, we first make a statistical analysis of the datasets. From Table~\ref{co-ob-table} we can see that for the five datasets, about 50\% points are observed in two images ($CO_i=2$). As $CO_i$ value increases, the percentage decreases. That is, the possibility of a point appearing in  a large number of images is low. This feature is exploited in our design to customize hardware design.

In our design Schur\_3 with two PEs, the first PE is customized for mainly processing points with $CO_i=2$. Due to the on-chip memory resource limitation of the target FPGA device, the second PE is designed with only two SPUs. Because the processing time of points with different $CO_i$ is different, the workloads of the two PEs need to be balanced. We design a software controller to assign workloads. For the five datasets, points with $CO_i \in [2, 10]$ are assigned to the first PE, while points with $CO_i \in [5, 50]$ are assigned to the second PE. Here the upper bound is set to 50 because the maximum $CO_i$ is 50 for datasets with up to 50 images.

\subsection{Resource usage}

\begin{table*}[t]
\caption{Resource utilization of hardware designs of the Schur elimination module. PE: Schur elimination processing element excluding matrix $\mathbf{S}$ storage. MEM\_S: matrix $\mathbf{S}$ storage. Others: input buffer, output buffer and accumulation unit.}
\begin{center}
\begin{tabular}{|c|c|c|c|c|c|}
\hline
\multicolumn{2}{|c|}{HW Designs}  &  FF & LUT & BRAM$^{\rm a}$ & DSP  \\
\cline{1-2}
 & Sub Module&(\%)&(\%)&(\%)&(\%)\\
\hline
             & PE        & 30151          &17913            & 41             &156        \\
\cline{2-6}
Schur\_1     & MEM\_S    & 17             & 497             & 84.5           &0      \\
\cline{2-6}
Double-precision& Others & 3889           & 2793            & 16             &20        \\
\cline{2-6}
              & Total    & 34057(22\%)    & 21203(27\%)     & 141.5(53\%)    &176(44\%)       \\
\hline
             & PE        & 11534          & 8626            & 21.5           &57        \\
\cline{2-6}
Schur\_1     & MEM\_S    & 16             & 250             & 43.5           &0      \\
\cline{2-6}
Single-precision & Others& 1286           & 1362            & 8              &9        \\
\cline{2-6}
              & Total    & 12836(8\%)     & 10238(13\%)     & 73(28\%)       &66(16\%)       \\
\hline
              & PE       & 13171          & 10650           & 28.5           &74        \\
\cline{2-6}
Schur\_2     & MEM\_S    & 28             & 462             & 85             &0        \\
\cline{2-6}
Single-precision& Others & 1506           & 1571            & 8              &11        \\
\cline{2-6}
              & Total    & 14705(9\%)     & 12683(16\%)     & 121.5(46\%)    & 85(21\%)       \\
\hline
              & PE\_Large& 13173          & 11103           & 28.5           & 69       \\
\cline{2-6}
              & PE\_Small& 13050          & 10656           & 15             & 74       \\
\cline{2-6}
 Schur\_3     & MEM\_S   & 56             & 921             & 169.5          & 0       \\
\cline{2-6}
Single-precision& Others & 2598           & 1854            & 16             & 21       \\
\cline{2-6}
             & Total     & 28877(18\%)    & 24534(31\%)     & 229(86\%)      & 164(41\%)      \\
\hline
\end{tabular}
\label{resource}
\end{center}
\footnotesize{$^{\rm a}$The number of Block RAMs is accounted in terms of 36Kbit block. 0.5 means a 18Kbit block.}\\
\end{table*}

\begin{table*}[t]
\caption{Performance comparison of hardware and software implementations of the Schur elimination over different datasets}
\begin{center}
\begin{tabular}{|c|c|c|c|c|c|c|c|}
\hline
{\textbf{Dataset}}&\multicolumn{5}{|c|}{\textbf{Execution time (ms)}}&\multicolumn{2}{|c|}{\textbf{FPGA Speedup}}\\
\cline{2-8}
  &\textbf{\textit{Schur\_1}}  &\textbf{\textit{Schur\_2}}  &\textbf{\textit{Schur\_3}}  &\textbf{\textit{Intel}} &\textbf{\textit{ARM}} &\textbf{\textit{Intel}} &\textbf{\textit{ARM}}\\
\hline
\textcircled{1}  &  59.155 & 31.746 & 15.885 &  43.168 & 817.381 & 2.717 & 51.45\\
\hline
\textcircled{2}  &  20.822 & 11.905 & 5.974  &  18.574 & 341.569 & 3.109 & 57.17\\
\hline
\textcircled{3}  &  50.261 & 27.560 & 13.854 &  35.607 & 662.382 & 2.570 & 44.93\\
\hline
\textcircled{4}  &  35.401 & 19.034 & 9.632  &  20.278 & 373.159 & 2.105 & 38.74\\
\hline
\textcircled{5}  &  66.847 & 36.010 & 18.126 &  43.707 & 805.415 & 2.411 & 44.43\\
\hline
Average          &  46.497 & 25.251 & 12.649 &  32.267 & 599.981 & 2.582 & 47.34\\
\hline
\end{tabular}
\label{TIME}
\end{center}
\end{table*}

The experimental results include three parts. The first part analyzes the characteristics of datasets, showing the distribution of $CO_i$. The second part shows the resource usage of the designs. The last part evaluates the speed and power consumption of the designs.

\subsection{Dataset analysis}

The resource usage of the three hardware designs of the Schur elimination module is reported in Table~\ref{resource}, including flip-flop (FF), lookup table (LUT), BRAM and DSP blocks. The resource consumption of important submodules is also shown. From the table we can make the following observations. Firstly, using single-precision floating point can save resources 14\% FFs, 14\% LUTs, 25\% BRAMs and 28\% DSPs, compared to double-precision floating point implementation, while the computation accuracy of the Schur elimination is not affected. Our experiment result shows that the difference of the Frobenius norm of matrix $\mathbf{S}$ is within $10^{-6}$ between double-precision and single-precision implementations. Secondly,
comparing Schur\_2 with Schur\_1, due to SPU duplication by 2, the computational resource usage increases up to 5\%, while the BRAM usage increases significantly 18\%. This is because two SPUs leads to almost doubling matrix $\mathbf{S}$ storage space. Thirdly, in Schur\_3, the two PEs are customized leading to DSP or BRAM saving. PE\_Small corresponds to the PE processing points with $2 \leq CO_i \leq 10$. This can save 13.5 BRAM blocks compared to the PE (for processing points with $2 \leq CO_i \leq 50$) in Schur\_2. PE\_Large corresponds to the PE processing points with $5 \leq CO_i \leq 50$. It saves 5 DSP blocks compared to the PE in Schur\_2. The DSP saving can be enlarged if the lower bound increases. Lastly, for the parallel Schur elimination implementation, on-chip BRAM is the main limit or obstacle. The consumption source mainly stems from matrix $\mathbf{S}$ storage. This also points out a future research direction on matrix $\mathbf{S}$ storage reduction.

\subsection{Execution time and power consumption}

We evaluate the execution time of Schur elimination with different datasets on different computing platforms. The results are shown in Table~\ref{TIME}. It is shown that on average Schur\_3 is 3.7 times and 2 times faster than Schur\_1 and Schur\_2, respectively. The hardware implementation Schur\_3 is about 3.1 times and 57 times faster than the Intel and ARM implementations, respectively.

Moreover, we also evaluate performance and power consumption of the BA implementations on the three computing platforms. Averaging over the five datesets, the execution time per BA iteration of Intel, ARM and our software-hardware design are 0.11s, 1.87s and 1.29s, respectively. The nominal power consumption of the used Intel CPU is 55W, while the power consumption of the ARM core and our design are 1.5W and 2.8W, respectively, reported by Xilinx power estimator. The Intel CPU implementation has the fastest speed, but is not suitable to embedded applications due to its high power consumption. Our design is 1.5 times faster than the ARM implementation. Note that, in our design the Schur elimination part is accelerated on hardware and the rest of BA is executed on the ARM core. The Schur elimination part takes about 30\% of the execution time of BA. In future, other parts of BA such as Jacobian update will also be accelerated on hardware to achieve higher performance improvement.

\section{Conclusion}
BA is a fundamental optimization technique used in many crucial applications, and often the primary performance and power consumption bottleneck in these applications. However, due to the complexities of BA algorithms, designing hardware to accelerate BA is extremely challenging. Previous approaches of optimizing BA performance heavily rely on parallel processing or distributed computing, which trade higher power consumption for higher performance. In this paper, we presented $\pi$-BA, the first hardware-software co-designed BA engine on an embedded FPGA-SoC. Specifically, we developed a novel Co-Observation Optimization technique, and experimental results confirmed that $\pi$-BA outperformed existing BA solutions in both performance and power consumption. With $\pi$-BA, we can enable more robotic localization as well as visual reconstruction applications by allowing larger scale online local BA on power-constrained embedded devices and more efficient offline global BA by using less computing resources and power consumption.

\bibliographystyle{IEEEtran}
\bibliography{IEEEabrv,references}

\begin{thebibliography}{10}
\providecommand{\url}[1]{#1}
\csname url@samestyle\endcsname
\providecommand{\newblock}{\relax}
\providecommand{\bibinfo}[2]{#2}
\providecommand{\BIBentrySTDinterwordspacing}{\spaceskip=0pt\relax}
\providecommand{\BIBentryALTinterwordstretchfactor}{4}
\providecommand{\BIBentryALTinterwordspacing}{\spaceskip=\fontdimen2\font plus
\BIBentryALTinterwordstretchfactor\fontdimen3\font minus
  \fontdimen4\font\relax}
\providecommand{\BIBforeignlanguage}[2]{{%
\expandafter\ifx\csname l@#1\endcsname\relax
\typeout{** WARNING: IEEEtran.bst: No hyphenation pattern has been}%
\typeout{** loaded for the language `#1'. Using the pattern for}%
\typeout{** the default language instead.}%
\else
\language=\csname l@#1\endcsname
\fi
#2}}
\providecommand{\BIBdecl}{\relax}
\BIBdecl

\bibitem{triggs1999bundle}
B.~Triggs, P.~F. McLauchlan, R.~I. Hartley, and A.~W. Fitzgibbon, \emph{Bundle
  adjustment—a modern synthesis}.\hskip 1em plus 0.5em minus 0.4em\relax
  Springer Berlin Heidelberg, 1999.

\bibitem{agarwal2010bundle}
S.~Agarwal, N.~Snavely, S.~M. Seitz, and R.~Szeliski, ``Bundle adjustment in
  the large,'' in \emph{European Conference on Computer Vision}, 2010, pp.
  29--42.

\bibitem{agarwal2009building}
S.~Agarwal, N.~Snavely, I.~Simon, S.~M. Seitz, and R.~Szeliski, ``Building
  {R}ome in a day,'' in \emph{Computer Vision, 2009 IEEE 12th International
  Conference on}.\hskip 1em plus 0.5em minus 0.4em\relax IEEE, 2009, pp.
  72--79.

\bibitem{mur2015orb}
R.~Mur-Artal, J.~M.~M. Montiel, and J.~D. Tardos, ``{ORB-SLAM}: a versatile and
  accurate monocular {SLAM} system,'' \emph{IEEE Transactions on Robotics},
  vol.~31, no.~5, pp. 1147--1163, 2015.

\bibitem{liu2017creating}
S.~Liu, L.~Li, J.~Tang, S.~Wu, and J.-L. Gaudiot, ``Creating {A}utonomous
  {V}ehicle {S}ystems,'' \emph{Synthesis Lectures on Computer Science}, vol.~6,
  no.~1, pp. i--186, 2017.

\bibitem{maimone2007two}
M.~Maimone, Y.~Cheng, and L.~Matthies, ``Two years of visual odometry on the
  {M}ars exploration rovers,'' \emph{Journal of Field Robotics}, vol.~24,
  no.~3, pp. 169--186, 2007.

\bibitem{klingner2013street}
B.~Klingner, D.~Martin, and J.~Roseborough, ``Street view
  motion-from-structure-from-motion,'' in \emph{Proceedings of the IEEE
  International Conference on Computer Vision}, 2013, pp. 953--960.

\bibitem{jeong2012pushing}
Y.~Jeong, D.~Nister, D.~Steedly, R.~Szeliski, and I.-S. Kweon, ``Pushing the
  envelope of modern methods for bundle adjustment,'' \emph{IEEE transactions
  on pattern analysis and machine intelligence}, vol.~34, no.~8, pp.
  1605--1617, 2012.

\bibitem{wu2011multicore}
C.~Wu, S.~Agarwal, B.~Curless, and S.~M. Seitz, ``Multicore bundle
  adjustment,'' in \emph{Computer Vision and Pattern Recognition}, 2011, pp.
  3057--3064.

\bibitem{eriksson2016consensus}
A.~Eriksson, J.~Bastian, T.-J. Chin, and M.~Isaksson, ``A consensus-based
  framework for distributed bundle adjustment,'' in \emph{Proceedings of the
  IEEE Conference on Computer Vision and Pattern Recognition}, 2016, pp.
  1754--1762.

\bibitem{zhang2017distributed}
R.~Zhang, S.~Zhu, T.~Fang, and L.~Quan, ``Distributed very large scale bundle
  adjustment by global camera consensus,'' in \emph{Proceedings of the IEEE
  International Conference on Computer Vision}, 2017, pp. 29--38.

\bibitem{strasdat2010real}
H.~Strasdat, J.~Montiel, and A.~J. Davison, ``{R}eal-time monocular {SLAM}:
  {W}hy filter?'' in \emph{Robotics and Automation (ICRA), 2010 IEEE
  International Conference on}.\hskip 1em plus 0.5em minus 0.4em\relax IEEE,
  2010, pp. 2657--2664.

\bibitem{Lourakis2005Is}
M.~Lourakis and A.~A. Argyros, ``Is {L}evenberg-{M}arquardt the most efficient
  optimization algorithm for implementing bundle adjustment?'' in
  \emph{Computer Vision, 2005. ICCV 2005. Tenth IEEE International Conference
  on}, vol.~2.\hskip 1em plus 0.5em minus 0.4em\relax IEEE, 2005, pp.
  1526--1531.

\bibitem{barrett1994templates}
R.~Barrett, M.~W. Berry, T.~F. Chan, J.~Demmel, J.~Donato, J.~Dongarra,
  V.~Eijkhout, R.~Pozo, C.~Romine, and H.~Van~der Vorst, \emph{Templates for
  the solution of linear systems: building blocks for iterative methods}.\hskip
  1em plus 0.5em minus 0.4em\relax Siam, 1994, vol.~43.

\bibitem{ceres-solver}
S.~Agarwal, K.~Mierle, and Others, ``Ceres {S}olver,''
  \url{http://ceres-solver.org}.

\end{thebibliography}
\end{document}